\documentclass[aps,prl,prabib,twocolumn,
showpacs,preprintnumbers,amsmath,amssymb,floatfix,aps]{revtex4}

\usepackage{graphicx}

\begin{document}

\title{Critical temperature of interacting Bose gases in two and three dimensions.
}
\author{S. Pilati and S. Giorgini}
\affiliation{Dipartimento di Fisica, Universit\`a di Trento and CNR-INFM BEC Center, I-38050 Povo, Trento, Italy}
\author{N. Prokof'ev}
\affiliation{Department of Physics, University of Massachusetts, Amherst, MA 01003, USA}
\affiliation{Theoretische Physik, ETH Z\"{u}rich, CH-8093 Z\"{u}rich, Switzerland}
\affiliation{Russian Research Center ``Kurchatov Institute'', 123182 Moscow, Russia}

\begin{abstract}
We calculate the superfluid transition temperature of homogeneous interacting Bose gases in three and two spatial dimensions using large-scale Path Integral Monte Carlo simulations (with up to $N=10^5$ particles). In 3D we investigate the limits of the universal critical behavior in terms of the scattering length alone by using different models for the interatomic potential. We find that this type of universality sets in at small values of the gas parameter $na^3 \lesssim 10^{-4}$. This value is different from the estimate $na^3 \lesssim 10^{-6}$ for the validity of the asymptotic expansion in the limit of vanishing $na^3$. In 2D we study the Berezinskii-Kosterlitz-Thouless transition of a gas with hard-core interactions. For this system we find good agreement with the classical lattice $|\psi|^4$ model up to very large densities. We also explain the origin of the existing discrepancy between previous studies of the same problem.
\end{abstract}

\maketitle

The theoretical determination of the superfluid transition temperature in homogeneous, interacting Bose systems is a fine example of a many-body problem that can be quantitatively addressed only by ``exact'' numerical techniques. This fact is well understood in the case of strongly interacting quantum fluids, such as liquid $^4$He where the critical temperature in bulk~\cite{PollockCeperley87} as well as in two-dimensional configurations~\cite{CeperleyPollock89,Boninsegni06} was calculated using path integral Monte Carlo (PIMC) simulations, but at first glance is surprising in the case of dilute gases. However, in three dimensions (3D) the presence of any finite interaction changes the universality class of the transition from the Gaussian complex-field model, corresponding to the ideal gas Bose-Einstein condensation (BEC) temperature $T_c^0$, to that of the XY model. Thus, the critical temperature $T_c$ can not be obtained from $T_c^0$ perturbatively~\cite{Baym99}. In two dimensions (2D) the superfluid transition, which belongs to the Berezinskii-Kosterlitz-Thouless (BKT) universality class~\cite{BKT1,BKT2}, is induced by interaction effects and there is no unperturbed critical temperature to start with.

In a 3D weakly repulsive gas the critical temperature shift is fixed by the $s$-wave scattering length $a$ ($a>0$), which characterizes interatomic interactions at low temperature~\cite{Gruter97,Baym99,HolzmannKrauth99,Kashurnikov01,ArnoldMoore01,Arnold01,Kastening03,NhoLandau04},
\begin{equation}
T_c=T_c^0\left[1+c(an^{1/3}) \right] \;.
\label{3DTc}
\end{equation}
Here $n$ is the gas density and $T_c^0=(2\pi\hbar^2/mk_B)[n/\zeta(3/2)]^{2/3}$ with $m$ the particle mass and $\zeta(3/2)\simeq2.612$. The numerical coefficient $c$ in Eq.~(\ref{3DTc}) was calculated in Refs.~\cite{Kashurnikov01,ArnoldMoore01} by solving the effective 3D classical $|\psi|^4$ model using lattice Monte Carlo simulations. The reported value is $c=1.29\pm0.05$. The same classical model was employed in Ref.~\cite{Arnold01} to calculate the higher-order logarithmic correction to   (\ref{3DTc}).

\begin{figure}
\begin{center}
\includegraphics[width=8.5cm]{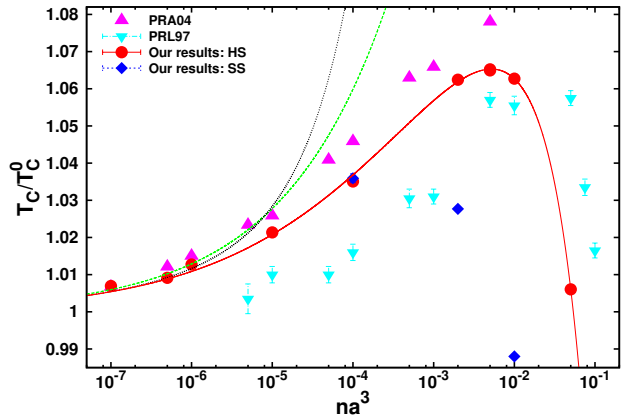}
\caption{(color online). Critical temperature in 3D as a function of the gas parameter $na^3$. The symbols labeled by PRA04 correspond to the results of Ref.~\cite{NhoLandau04}, the ones labeled by PRL97 correspond to Ref.~\cite{Gruter97}. The dashed line (green) is the expansion (\ref{3DTc}) of Ref.~\cite{Kashurnikov01} and the dotted line (black) is the expansion of Ref.~\cite{Arnold01} including logarithmic corrections. The solid line (red) is a guide to the eye.}
\label{fig1}
\end{center}
\end{figure}

Continuous-space studies of a gas of hard spheres, based on the conventional PIMC algorithm~\cite{Ceperley95}, were carried out in Refs.~\cite{Gruter97,NhoLandau04}.
Both calculations suffered from two shortcomings: first, the number of particles in the simulation was only few hundreds making the extrapolation to the thermodynamic limit difficult; second, the algorithm is known to be inefficient for
simulations of the superfluid density. The results of Ref.~\cite{NhoLandau04} for $T_c$ agree with the asymptotic law (\ref{3DTc}) at small densities, in contrast to the significantly lower values obtained in Ref.~\cite{Gruter97} (see Fig.~\ref{fig1}). Moreover, there is strong (about ten standard deviations!) discrepancy between Refs.~\cite{Gruter97} and \cite{NhoLandau04} at higher densities calling for further investigation of the problem.

In 2D the BKT transition temperature of a weakly interacting gas is written in the form~\cite{Popov87}
\begin{equation}
T_\text{BKT}=\frac{2\pi\hbar^2n_{2D}}{mk_B}\frac{1}{\ln(\xi/4\pi)+\ln\ln(1/n_{2D}a_{2D}^2)} \;,
\label{2DTc}
\end{equation}
where $n_{2D}a_{2D}^2$ is the corresponding gas parameter, and $a_{2D}$ is the 2D scattering length. The numerical coefficient $\xi$ was calculated in Ref.~\cite{Prokofev01} from lattice Monte Carlo simulations of the classical $|\psi|^4$ model, similarly to the 3D case, yielding the value $\xi=380\pm3$.

An important question concerns the range of applicability of Eqs.~(\ref{3DTc})-(\ref{2DTc}) since they were derived for the limit of vanishingly small interaction strength. More broadly, one is interested in knowing up to what value of the gas parameter it is possible to express $T_c$ as a function of $na^3$ alone and ignore the dependence on the interaction potential details. These questions are particularly relevant for the 2D case where experimental determinations of the critical temperature in trapped configurations have become available~\cite{Hadzibabic06,Kruger07}.

In this Letter we report PIMC results for the superfluid transition temperature in interacting Bose gases both in 3D and in 2D. We carry out large-scale simulations of homogeneous systems with up to $N=10^5$ particles. The simulations are based on the worm algorithm, recently extended to continous-space systems, which allows for a reliable and efficient calculation of the superfluid density~\cite{Boninsegni06}.
With new data the extrapolation of the critical temperature to the thermodynamic limit can be done with the level of accuracy that was unreachable in previous attempts. In 3D we determine the dependence of $T_c$ on the gas parameter $na^3$, reporting good agreement with the expansion (\ref{3DTc}) in the dilute regime and significant deviations from previous studies~\cite{Gruter97,NhoLandau04} at higher densities. Furthermore, we carry out simulations with both a hard- and a soft-sphere interatomic potential to investigate the universal behavior in terms of the scattering length. In 2D we calculate $T_\text{BKT}$ for a hard-disk gas as a function of the interaction strength, finding results in excellent agreement with the prediction (\ref{2DTc}) up to regimes of surprisingly high density.

We consider a 3D system of $N$ particles with periodic boundary conditions described by the Hamiltonian $H=-(\hbar^2/2m)\sum_{i=1}^N\nabla_i^2+\sum_{i<j}V(|{\bf r}_i-{\bf r}_j|)$, where ${\bf r}_i$ denotes the coordinates of the $i$-th particle. Two-body interactions are modeled by the following potentials: the hard-sphere (HS) potential, $V^{HS}(r)=+\infty$ if $r<a$ and zero otherwise, and the soft-sphere (SS) potential, $V^{SS}(r)=V_0$ ($V_0>0$) if $r<R_0$ and zero otherwise. In the HS case the $s$-wave scattering length $a$ coincides with the range of the potential, while in the SS case it is given by $a=R_0[1-\tanh(K_0R_0)/K_0R_0]$ with $K_0=\sqrt{V_0m}/\hbar$. We always use the range $R_0=5a$ and adjust $V_0$ to obtain the desired value of $a$. We notice that the HS and SS model represent two extreme cases of repulsive potentials: in the HS case the energy is entirely kinetic, while for the SS case the Born approximation result $a_B=(m/\hbar^2)\int_0^\infty V(r)r^2dr$ accounts for more than 80\% of the value of the scattering length.

\begin{figure}
\begin{center}
\includegraphics[width=8.5cm]{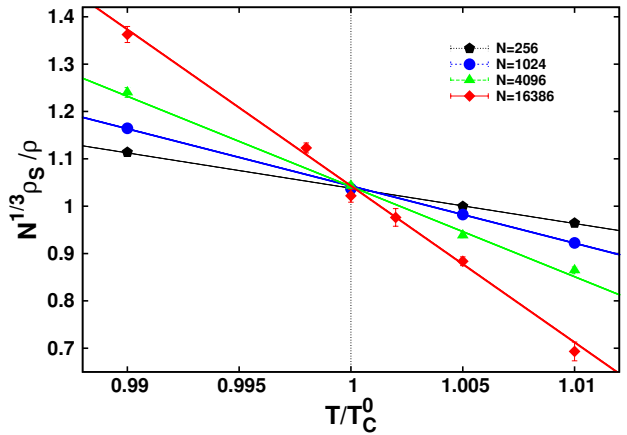}
\caption{(color online). Results for a 3D non-interacting gas. Scaled superfluid fraction as a function of temperature for different system sizes. The lines are linear fits from Eq.~(\ref{scaling}).}
\label{fig2}
\end{center}
\end{figure}

In a PIMC simulation one obtains averages of physical quantities over a set of stochastically generated configurations ${\bf R}=({\bf r}_1,...,{\bf r}_N)$ sampled from a probability distribution proportional to the density matrix $\rho({\bf R},{\bf R},\beta)=\langle{\bf R}|e^{-\beta H}|{\bf R}\rangle$, where $\beta=1/k_BT$ is the inverse temperature. The superfluid density $\rho_s$ is obtained from the winding number estimator~\cite{PollockCeperley87}, which accounts for long permutation cycles of identical particles occurring in the system. The calculation of $\rho({\bf R},{\bf R},\beta)$ is based on the pair-product decomposition, where the two-body density matrix associated with the relative motion of the pair is determined exactly both for the HS and the SS potential~\cite{Pilati06,Pilati07}. The critical temperature $T_c$ is determined from calculations of the superfluid fraction $\rho_s/\rho$ ($\rho=mn$ is the total mass density) for systems with increasing particle number $N$ using the scaling ansatz\cite{PollockRunge92}
\begin{equation}
N^{1/3}\rho_s(t,N)/\rho=f(tN^{1/3\nu})= f(0)+f^\prime(0)tN^{1/3\nu}+... \;.
\label{scaling}
\end{equation}
Here, $t=(T-T_c)/T_c$ is the reduced temperature, $\nu$ is the critical exponent
of the correlation length $\xi(t)\sim t^{-\nu}$, and $f(x)$ is a universal analytic function which allows for a linear expansion around $x=0$.

\begin{figure}
\begin{center}
\includegraphics[width=8.5cm]{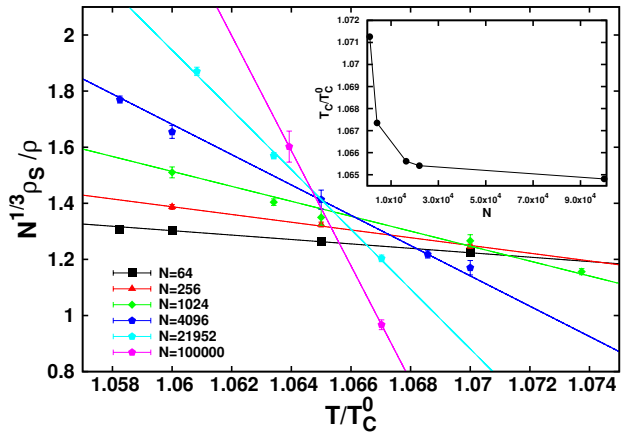}
\caption{(color online). Results for a 3D hard-sphere gas with $na^3=5\times10^{-3}$. Scaled superfluid fraction as a function of temperature for different system sizes. The lines are linear fits from Eq.~(\ref{scaling}). The inset shows the $N$ dependence of the intersection point between lines corresponding to pairs of consecutive system sizes.}
\label{fig3}
\end{center}
\end{figure}

We first consider the non-interacting case (see Fig.~\ref{fig2}). The scaling curves all intersect at the same value of the reduced temperature according to Eq.~(\ref{scaling}) with $T_c/T_c^0=1.0005(4)$ and $\nu=0.96(3)$. The value of $\nu$ is consistent with the prediction $\nu=1$ of the complex Gaussian model. The corresponding results for the hard-sphere gas at the density $na^3=5\times 10^{-3}$ are reported in Fig.~\ref{fig3}. In this case the intersection point between curves for consecutive system sizes clearly drifts towards lower temperatures for larger $N$ (see inset of Fig.~\ref{fig3}). This effect arises from corrections to the scaling law [the right hand side of Eq.~(\ref{scaling}) has to be multiplied by $(1+ CN^{-\omega/3}+\dots )$ with $\omega \approx 0.8$ and $C$ of order unity]. A reliable extrapolation to the thermodynamic limit using Eq.~(\ref{scaling}) requires large-scale systems on order of $N\gtrsim10^4$. The temperature $T_c$ and the exponent $\nu$ are determined from Eq.~(\ref{scaling}) by considering series of data corresponding to systems with $N$ sufficiently large. We find $T_c/T_c^0=1.0652(4)$ (see Table~\ref{tab1}) and $\nu=0.70(4)$, in agreement with the value $\nu=0.67$ corresponding to the universality class of the XY model in three dimensions~\cite{GuillouZinnJustin77}. By using the same procedure we calculate $T_c$ as a function of the gas parameter $na^3$ both for the HS and the SS potential. The results are shown in Fig.~\ref{fig1} and are reported in Table~\ref{tab1}.

Except for the largest densities, our results for the HS gas are systematically higher than the ones of Ref.~\cite{Gruter97} and lower compared to the ones of Ref.~\cite{NhoLandau04}~\cite{note}. With much better accuracy for $\rho_s$ and larger system sizes we are in the position to explain the origin of discrepancy.
It is two-fold. First, the number of imaginary time slices used in Ref.~\cite{NhoLandau04} was 15, a factor of 3 larger than in
Ref.~\cite{Gruter97}. We find, by doing simulations with
up to 96 slices, that about 25 slices have to be used to ensure that
the corresponding systematic errors are negligible. Second, Refs.~\cite{Gruter97,NhoLandau04} underestimated error bars by a factor of ten because
multiple intersections between scaling curves (i) render the procedure of locating
the intersection point ambiguous, and (ii) prevent one from detecting corrections to scaling and the flow of intersection points with the system size (which is substantial between $N=256$ and $N\sim 20000$).
The discrepancy with Ref.~\cite{NhoLandau04} reduces at very small densities $na^3 \lesssim 10^{-6}$, where finite-size corrections to $T_c$ appear to be less relevant and there is agreement with the expansion (\ref{3DTc}). The critical temperature $T_c$ first increases with $na^3$, then goes through a maximum and for larger values of the gas parameter decreases below the $T_c^0$ value. For example, $T_c/T_c^0=0.70$ in liquid $^4$He corresponding to the effective gas parameter $na^3\simeq0.21$~\cite{Kalos74}. The HS gas is finally expected to become a solid at the freezing density $na^3\simeq0.25$~\cite{Kalos74}. Concerning the comparison between different model potentials, at $na^3=10^{-4}$ we find very good agreement between the HS and SS gas, while for higher densities deviations start to become evident and the SS results are significantly smaller.

\begin{table}
\centering
\begin{ruledtabular}
\caption{Transition temperature $T_c$ of a 3D hard-sphere gas for different values of the gas parameter. The results with the label SS correspond to a soft-sphere gas.}
\begin{tabular}{cccc}
$na^3$  & $T_c/T_c^0$ & $na^3$ &  $T_c/T_c^0$ \\
\hline
$1\times10^{-7}$      &  1.0069(10)  &  $2\times10^{-3}$       &  1.0624(4)  \\
$5\times10^{-7}$      &  1.0091(7)   &  (SS) $2\times10^{-3}$  &  1.0277(4)  \\
$1\times10^{-6}$      &  1.0127(7)   &  $5\times10^{-3}$       &  1.0652(4)  \\
$1\times10^{-5}$      &  1.0214(7)   &  $1\times10^{-2}$       &  1.0627(4)  \\
$1\times10^{-4}$      &  1.0351(7)   &  (SS) $1\times10^{-2}$  &  0.9880(5)  \\
(SS) $1\times10^{-4}$ &  1.0359(12)  &  $5\times10^{-2}$       &  1.0060(5)  \\
\label{tab1}
\end{tabular}
\end{ruledtabular}
\end{table}

\begin{figure}
\begin{center}
\includegraphics[width=8.5cm]{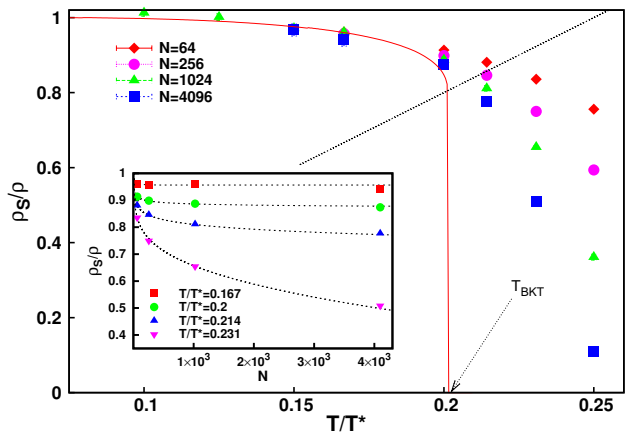}
\caption{(color online). Results for a 2D hard-disk gas with $na^2=0.01$. Superfluid density as a function of temperature for different system sizes. The solid line (red) shows the extrapolation to the thermodynamic limit. The dashed line (black) corresponds to the BKT universal jump. In the inset we show the dependence of the superfluid density on the system size for different temperatures. The dotted lines are fits using the Kosterlitz-Thouless renormalization group equations.}
\label{fig4}
\end{center}
\end{figure}

\begin{figure}
\begin{center}
\includegraphics[width=8.5cm]{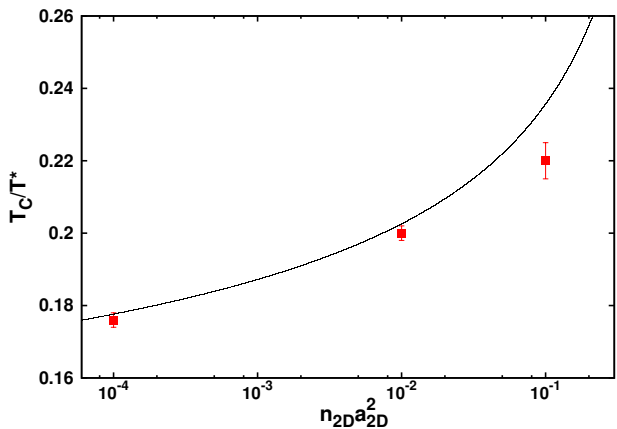}
\caption{(color online). Critical temperature of a 2D gas of hard disks as a function of the gas parameter $na^2$. The solid line is the result (\ref{2DTc}).}
\label{fig5}
\end{center}
\end{figure}

The simulations of the 2D Bose gas are carried out in a similar way, using a hard-core potential with range $a_{2D}$ (hard-disk potential) to model the interactions. The results for the superfluid fraction as a function of temperature are reported in Fig.~\ref{fig4} for the value $n_{2D}a_{2D}^2=0.01$ of the gas parameter. The extrapolation to the thermodynamic limit is carried out by fitting numerical data at finite $N$ to the solution of the Kosterlitz-Thouless recursion relations\cite{BKT2}
\begin{eqnarray}
df(N)/d\ln N&=&-y^2(N)f^2(N)/2
\nonumber\\
dy(N)/d\ln N&=&[1-f(N)]y(N) \;,
\label{KT}
\end{eqnarray}
where $f(N)=(\hbar^2\pi n_{2D}/2mk_BT)\rho_s(N)/\rho$ is a dimensionless function proportional to the superfluid fraction and $y(N)$ is proportional to the vortex fugacity. The starting values $f_0$ and $y_0$ of the recursion relations (\ref{KT}) are determined from a best fit to the results corresponding to different values of $N$ and system temperature in close proximity of the transition (see Fig.~\ref{fig4}). From these initial values one determines the critical temperature $T_\text{BKT}$ in the thermodynamic limit. Here $T_\text{BKT}$ is written in units of $T^\ast=2\pi\hbar^2n_{2D}/(2mk_B)$, which provides the natural temperature scale for quantum degeneracy in 2D. In Fig.~\ref{fig4} we also show the prediction of the universal jump of the superfluid fraction at the transition $\rho_s/\rho=2mk_BT_\text{BKT}/(\pi\hbar^2n_{2D})$~\cite{NelsonKosterlitz77}, in nice agreement with the temperature dependence of our extrapolated curve. The results for $T_\text{BKT}$ as a function of $n_{2D}a_{2D}^2$ are reported in Fig.~\ref{fig5}, where they are compared with the prediction obtained from Eq.~(\ref{2DTc}). The agreement is surprisingly good up to very large values of the gas parameter. At the highest density $n_{2D}a_{2D}^2=0.1$, which is close to the freezing point $n_{2D}a_{2D}^2\simeq0.32$ of a gas of hard disks~\cite{Xing90}, the dilute gas result (\ref{2DTc}) is only about 7\% larger than the PIMC value.

In conclusion, we have carried out a numerical study of the superfluid transition temperature of interacting Bose gases in 3D and 2D and established the limits of validity of the asymptotic expansions (\ref{3DTc}) and (\ref{2DTc}), and the universal description in terms of the scattering length. We have also explained and resolved the
discrepancy between previous studies.


\begin{thebibliography}{99}
\bibitem{note} We notice that by using small system sizes similar to the ones of Ref.~\cite{NhoLandau04}, the value of $T_c$ increases becoming consistent with the findings of that paper.
\end{thebibliography}
\end{document}